# Self-gravitational red shift effect and micro-black holes production in dipolar electromagnetic sources

**M Auci**
Odisseo Space, via Battistotti Sassi 13, 20133 Milano, Italy

E-mail: massimo.auci@odisseospace.it

**Abstract.** Using Bridge Theory to describe the electromagnetic interactions occurring between high energy pairs of particles, we predict an anomalous self-gravitational red-shift in the frequency of the EM source produced during the interaction. The extreme consequence is the introduction of an upper limit in the electromagnetic spectrum. Hypothesizing a scattering energy greater than $5.55\ 10^{22}$ Mev, the source collapses under his events horizon, becoming a neutral spin less micro-black hole: perfect candidates to contribute to the dark matter of the Universe.

## 1. Introduction

The Bridge Theory (BT) approach[1] to unifying Classical and Quantum Electrodynamics is been proposed in the 1999. The theory is based on the role of the transversal component of the Poynting vector[2-4] to localise in the neighbourhood of a dipolar electromagnetic source (DEMS) an amount of energy and momentum in agreement with the quantum predictions for a photon.

A DEMS is originated during an electromagnetic collision between a real or virtual pair of particles. For the total energy and momentum conservation laws the energy and momentum exchanged by particles during the collision decide the frequency and the wavelength of the electromagnetic (EM) source: from the Quantum Theory point of view we have that the characteristic wavelength and frequency of the DEMS is that one of the photon responsible of the energy and momentum exchange between the interacting particles, i.e. the DEMS coincides physically with the exchanged photon.

Since the theoretical evaluation of Planck's constant resulting from BT is totally in agreement with its experimental value, when we use the symbol $\hbar$ we assume the usual numerical value of the constant but non the quantum theory formalism, because in BT the ordinary quantum phenomenology is substituted self-consistently by the like-quantum behaviour of the DEMS. These considerations leads to the assumption of a total photon-DEMS equivalency, authorising us to analyse the energy limits inside of the DEMS. A peculiar behaviour of the DEMS is that to produce a sort of inertia when energy is localised in the neighbourhood of the source. The inertia inside the DEMS induces a self-gravitational distortion in the EM field structure of the source, i.e. of the photon, when the collision energy between the particles grows. This introduces a cut in the total energy that a particle can acquire during the creation or during its interaction with others particles but could be also the cause of gravitational collapse of the entire DEMS.

## 2. DEMS creation and inertial effect in the source zone expansion

Following BT we will consider the dynamical behaviour of the source zone[3] (SZ) during the EM interaction of two oppositely charged particles.

The EM interaction starts when each charge feels the EM field produced by the respective anti-charge. The SZ is proved be a spherical crown of growing thickness (initially zero), expanding[5] with velocity:

$$v_{\exp} = \frac{c}{\omega t} \tag{1}$$

decreasing with the elapsed time $t$ from the initial instant.

Since the time $t$ depends on the radius $r$ of the actual external spherical bound $\Sigma_r$ of the SZ, where $r$ grows taking values close to the production zone $\lambda/2 \le r \le 3\lambda/4$ of the DEMS, we define:

$$t = \frac{2(r - \lambda/2)}{c} \tag{2}$$

as the elapsed time from the beginning of the EM interaction. When the radius $r$ varies, the elapsed time takes values within the interval $0 \le t \le T/2$. Simultaneously occurs that: (a) the expansion velocity (1) of the spherical surface $\Sigma_r$ bounding the actual source zone decreases, passing from infinite when $r = \lambda/2$ (starting surface delimiting the internal side of the SZ: $\Sigma_0$) to $c/\pi$ when $r = 3\lambda/4$ (ending surface delimiting the external side of the SZ: $\Sigma_{max}$); (b) during the activity of the source occurring in the collision time T/2, not all the energy localised is instantaneously carried away by the radial component of the Poynting vector, because a growing fraction associated to the non zero transversal component is progressively stored inside the SZ with the effect of to increase gradually the EM mass of the source up to the maximum value $\hbar\omega/c^2$ characteristic of the DEMS.

The correlation of the effects (a) and (b) suggests that the decreasing of the velocity $v_{\exp}$ is due to an inertia[6] growing in time inside the SZ.

## 3. Mass density and effective active mass

Considering a single DEMS, its energy and momentum localised by the transversal component of the Poynting vector[1-4] $S_t$ are both supplied by the total energy and momentum of two colliding particles. Since the maximum local momentum[7] produced by the expansion of the external spherical surface is given by

$$\delta p_{\exp} = \frac{2\pi}{c^2} \int_{V_{\delta\Sigma}} S_t d^3x \ , \tag{3}$$

when the SZ gets its maximum extension, $\Sigma_{max}$ has expansion velocity $v_{\exp} = c/\pi$. Using momentum (3), we define in the same dynamical conditions the local EM energy contribution due to the expansion kinetic energy of the source as defined in Refs. 7, 8:

$$\mathrm{I}_{\exp} = \frac{1}{2} \delta p_{\exp} v_{\exp} = \frac{1}{c} \int_{V_{\delta\Sigma}} S_t d^3x \, . \tag{4}$$

In general, the contribution (4) can be considered obtained by the momentum (3), which pure EM mass value results by the extension achieved by the SZ, and by the expansion velocity (1) at time $t$. In this sense Eq. (4) is the energy contribution achieved by the DEMS when the SZ expansion ends; it follows that for time lower than T/2, Eq. (4) is depending by the EM mass and by the square of the expansion velocity achieved at this time. These last considerations, allow to an observer placed externally to the SZ to fill energy and momentum without internal time structure, i.e. as phenomena independent from the effective evolution of the SZ. In this case the observer measures the energy and momentum of a DEMS as time-independent values.

From the point of view of the mass energy localised inside the SZ, the phenomenology is not the same described above. In fact the mass is localised starting to zero along all the SZ expansion, so during the source formation, the observer fills the local mass density as due to all the EM energy produced during the life of the source, i.e. the DEMS mass has a time-distribution.

Using eq. (4), we define the contribution to the kinetic energy density localised in the SZ during the expansion process:

$$\delta\Psi = \left(\frac{d\text{l}_{\exp}}{dV}\right)_{V_{\exp}} = \frac{1}{2}\delta\left(\frac{dp}{dV}\right)_{V_{\exp}} v_{\exp}$$
$$= \frac{1}{2}\delta\mu\, v_{\exp}^2 = \frac{1}{c}S_t = \frac{1}{4\pi}\frac{e^2}{r^4}\Theta_t(\theta)\ . \qquad (5)$$

Integrating Eq. (5) over all angles of emission and using the definition of the spin action constant as in Refs. 9-11, the total kinetic energy density due to the expansion of the SZ of the DEMS turns out to be,

$$\Psi = \frac{1}{2}\mu\, v_{\exp}^2 = \oint\int_0^{\pi} \delta\Psi d\theta\, \mathrm{d}\varphi = \frac{3c}{8\pi}\frac{\mathrm{H}_{spin}}{r^4}\ . \qquad (6)$$

In order to measure the effective EM mass density, condensed during the SZ time-evolution, occurs to consider the expansion velocity deriving from Eqs. (1) and (2). According to the Eq. (6), let $\chi = 2r/\lambda$ be a suitable dimensionless variable varying in the SZ within the interval

$$1 \le \chi \le 3/2\ ,$$

and $r = \pi\, c\chi/\omega$ be the consequent expression of the radius of the spherical wave front, the mass density will then be

$$\mu = \frac{2\Psi}{v_{\exp}^2} = \frac{3}{\pi^3}\mathrm{H}_{spin}\frac{\omega^4}{c^5}\frac{(\chi-1)^2}{\chi^4}\ . \qquad (7)$$

By integrating the Eq. (7) over the volume of the spherical crown within the SZ interval, where the element

$$d^3x = v_{\exp}\, dt\, d\sigma = \frac{\pi^2 c^3}{\omega^3}\frac{\chi^2}{\chi-1}d\chi \sin\theta\, d\theta\, d\varphi \qquad (8.1)$$

is the volume of the infinitesimal portion of the expanding crown of thickness $v_{\exp}dt$ in the new variable $\chi$, we obtain the effective acting mass:

$$\mathrm{M}_{ef} = \int_{V_{\delta\Sigma}} \mu\, d^3x = 12\mathrm{H}_{spin}\frac{\omega}{c^2}\int_1^{3/2}\frac{\chi-1}{\chi^2}d\chi$$
$$= \left(12\ln\frac{3}{2} - 4\right)\mathrm{H}_{spin}\frac{\omega}{c^2} \qquad (8.2)$$

Since just the transversal component of the Pointing vector contributes to the effective EM mass $\mathrm{M}_{ef}$, only this component is responsible for the energy of the SZ and for the consequent increase of mass density within the spherical crown $\delta\Sigma$. Therefore the value of the action constant in Eq. (8b) does not include the mechanical energy[11-12] produced during the collision. In any case the action constants $\mathrm{H}_{spin}$ and $\hbar$ are numerically very close, hence we can approximate eq. (8) using $\hbar$, as made in Ref. 13, by writing

$$\mathrm{M}_{ef} \cong \left(12\ln\frac{3}{2} - 4\right)\frac{\hbar\omega}{c^2} \qquad (9)$$

where the numerical factor in eq. (9)

$$f = \frac{\mathrm{M}_{ef}c^2}{\hbar\omega} \cong \left(12\ln\frac{3}{2} - 4\right) \cong 86.56\,\% \qquad (10)$$

is the DEMS energy fraction reserved to the mass energy of the emerging particles.

Considering now the self-gravity produced by the mass inside the SZ, we suggest that the self-gravitational interaction is proportional to the increase of mass along the time evolution. In fact the equation (10) points out that not all the DEMS (i.e. photon) energy can act gravitationally on own SZ, because the EM energy localised inside the source is not a constant during the life time of the DEMS, but grows achieving the maximum value $E_s = \hbar\omega$ when $t = T/2$. This means that during the expansion, the energy acting as effective mass is only a growing fraction of the final total energy, i.e. respect to the time zero, the gravitational interaction is proportional to the mass varying with the time; so the density corresponds roughly to a mean over the time interval [0, *T* / 2].

Since *f* represents the mass fraction that characterises all the expansion process of the SZ, we can break up the total EM energy of the source into two terms:

$$E_s = (1-f)E_s + fE_s = E_{\bar{G}} + E_G \quad . \tag{11}$$

The term $E_G = fE_s$ is the only amount of the energy of the source, acting self-gravitationally on the SZ.

## 4. Self-gravitational potential, field and energy of a DEMS

Above we have assumed that in the DEMS the localised mass acts self-gravitationally. From classical definition of gravitational potential:

$$\Phi = -\gamma \int_{V_{\delta\Sigma}} \frac{\mu}{R} d^3x \ , \tag{12}$$

where γ is the gravitational constant, we will obtain the potential produced by the effective active mass $M_{ef}$ inside the expanding SZ.

From Eq. (7), let

$$\mu \cong \frac{3}{\pi^3} \hbar \frac{\omega^4}{c^5} \frac{(\chi-1)^2}{\chi^4} \tag{13}$$

be the mass density corrected for the mechanical action term and

$$d^3x = v_{\exp}\, dt\, d\sigma = \frac{c}{\omega} \frac{R}{\chi - 1} dR\, d\chi\, d\varphi \tag{14}$$

be the volume of an infinitesimal zone of the expanding crown $\delta\Sigma$, obtained using Eq. (8a), where R is the distance varying within the interval $0 \le R \le \lambda\chi$ between two points P and P' on the same expanding spherical surface $\Sigma_r$ (see Ref. 14) and $\varphi$ the azimuth varying within the interval $0 \le \varphi \le 2\pi$, by integration we get

$$\begin{aligned} \Phi &\cong -\frac{6}{\pi} \gamma \frac{\hbar}{c^3} \omega^2 \left(1 - \frac{1}{\chi}\right)^2 \\ &= -24\pi \frac{E_s^2}{E_P^2} \left(1 - \frac{1}{\chi}\right)^2 c^2 \end{aligned} \tag{15}$$

where $E_P = 2\pi \mathsf{E}_P = 7.67\,10^{22}$ Mev is the DEMS energy at Planck's length and

$$\mathsf{E}_P = \sqrt{\frac{\hbar c^5}{\gamma}} = m_P c^2 \tag{16}$$

is the mass energy corresponding to Planck's mass $m_P$.

We would like to point out that if γ and $\hbar$ not vary too much for effect of hidden variables, Eqs. (15) and (16) introduce a mass scale in which Planck's mass has a primary role.

In order to calculate the work done by the gravitational field during the expansion of the SZ, it is necessary to calculate the difference between the potential energies of the source when its external surface coincides respectively with the start surfaces $\Sigma_0$ and with one of the growing surface $\Sigma_r$.

Since

$$U_r = \frac{1}{2}\int \mu\, \Phi\, d^3x$$
$$= -24\pi \frac{E_s^3}{E_P^2}\left(6\ln\chi + \frac{18}{\chi} - \frac{9}{\chi^2} + \frac{2}{\chi^3} - 11\right) \qquad (17)$$

is the total potential energy of the source when the external surface has achieved the radius $r = \lambda\chi/2$, the work done during the expansion can be written as

$$E_g = U_0 - U_r \; . \qquad (18)$$

From the Eq. (17) results that at $t = 0$, i.e. $\chi = 1$, the potential energy $U_0$ is zero, now using Eq. (18) we calculate the total gravitational energy of the SZ for $\chi = 3/2$ to be

$$E_g \cong \Gamma \frac{E_s^3}{E_P^2} \qquad (19)$$

where

$$\Gamma = 144\pi\left(\ln\frac{3}{2} - \frac{65}{162}\right) \cong 1.91$$

is a dimensionless constant.

## 5. Electromagnetic emission

Because of the gravitational term (19), the total localised energy of the SZ $E_s$ is subject to a reduction as a consequence of a self-gravitational interaction. In fact when the SZ achieves the maximum space extension, the available localised EM energy for the emission is lower than the energy produced,

$$\mathsf{E} = E_s - E_g = \left(1 - \Gamma\frac{E_s^2}{E_P^2}\right)E_s \le E_s \; . \qquad (20)$$

If we consider an electron-positron interaction at low energy, scattering or pair annihilation can occur and in these cases the DEMS produced cannot have a total energy lower than two electron masses ($E_s \ge 1.02$ Mev). In both cases, using Eq. (20) the energies ratio

$$\eta = \frac{\mathsf{E}}{E_s} = \left(1 - \Gamma\frac{E_s^2}{E_P^2}\right) \qquad (21)$$

represents the output, i.e. the fraction of the available energy reserved for the source emission including both radiative and massive contributions.

According to Eq. (10), not all the energy of the source is available for the mass production. If the energy reserved to mass production is lower than two electron masses:
$E_G = f\, E_s < 1.02$ Mev i.e. $E_s < 1.18$ Mev, only EM primary reemission can occur with total energy equal to the available energy of the DEMS (20). Here $\eta$ represents the EM output of the DEMS. Eq. (21) shows that the output collapses to zero when the energy of the source becomes equal to
$E_P/\sqrt{\Gamma} \cong 5.55\ 10^{22}$ Mev.

Let $E_{col} = E_P/\sqrt{\Gamma}$ be the energy of the output collapse, using Eq. (21) we can write that the output of the DEMS in the entire energy range $0 \le E_s \le E_{col}$ is

$$\eta = 1 - \frac{E_s^2}{E_{col}^2} \; . \qquad (22)$$

## 6. Spectrum analyse and ways of emission in the model

The energy $E_s$ of a DEMS, besides characterising the available total energy (see eq. (20)), decides the most probable way of emission. In fact, when the energy $E_s$ increases, the output of the source (22) decreases, with the effect of reducing the maximum value of energy reserved for pure EM emission:

$$\mathcal{E}_{EM} = \mathcal{E} - E_G \tag{23}$$

We would like to comment in more detail upon this peculiar aspect. Eq. (23) shows that the available energy for EM emission depends on the difference between the amounts of available total energy

$$\mathcal{E} = \eta\, \hbar\omega \tag{24}$$

and the energy reserved for mass production

$$E_G = f\, \hbar\omega \tag{25}$$

When the total available energy is equal to the energy (25): $\mathcal{E} = E_G$, i.e. $\eta = f$, no primary EM emission (23) can occur, $\mathcal{E}_{EM} = 0$. In this case the DEMS can emit energy just as mass. We define the threshold energy related to the previous condition as the EM *limit*

$$E_{\lim} = \sqrt{1-f}\, E_{col} = 2.03\ 10^{22}\ \text{Mev}, \tag{26}$$

this value corresponds to an available mass energy

$$\mathcal{E}_{\lim} = \eta E_{\lim} = 1.76\ 10^{22}\ \text{Mev}. \tag{27}$$

For values of the DEMS energy $E_s$ bigger than the upper value, but within the interval

$$2.03\ 10^{22}\ \text{Mev} < E_s \le 3.2\ 10^{22}\ \text{Mev}, \tag{28}$$

the mass energy continues to grow up to until it achieves the maximum of the available mass of the spectrum for $E_{crit} = 3.2\ 10^{22}$ Mev

$$\mathcal{E}_{\max} = \eta E_{crit} = 2.14\ 10^{22}\ \text{Mev}. \tag{29}$$

For energies bigger than the critical value the massive emission decreases and it collapses for $E_s = E_{col}$.

For energies of the DEMS lower than $E_{\lim}$, the EM emission can occur alone or along with a massive emission. Using eqs. (23)-(25), we obtain that the part corresponding to a pure EM emission varies as a function of the energy $E_s$ as

$$\mathcal{E}_{EM} = \left(1 - f - \frac{E_s^2}{E_{col}^2}\right) E_s. \tag{30}$$

Since the condition $\frac{d\mathcal{E}_{EM}}{dE_s} = 0$ decides the value of $E_s$ corresponding to the maximum of EM emission, we obtain that, when the DEMS has energy

$$E_{EM-\max} = \sqrt{\frac{1-f}{3}}\, E_{col} = 1.17\ 10^{22}\ \text{Mev}\ , \tag{31}$$

the EM emission is maximal:

$$\mathcal{E}_{EM-\max} = \frac{2}{3}(1-f)\, E_{EM-\max} = 1.05\ 10^{21}\ \text{Mev}. \tag{32}$$

Since the maximum of EM emission is achieved when the DEMS energy is $1.17\ 10^{22}$ Mev, the characteristic temperature of an emitting black-body cannot exceed the value of energy corresponding to the EM maximum: $1.05\ 10^{21}$ Mev. This means that $10^{21}$ Mev is the energy cut of the EM spectrum.

In the model the existence of the self-gravitational red shift in the EM interactions and of an upper limit in the EM spectrum imposes a thermal bound in the Universe evolution. In fact for each DEMS produced in the past, present and future, a small part of the engaged energy is gravitationally degraded with the effect of to reduce progressively the local temperature without that other processes can warm the local Universe with an enough efficient local mechanism.

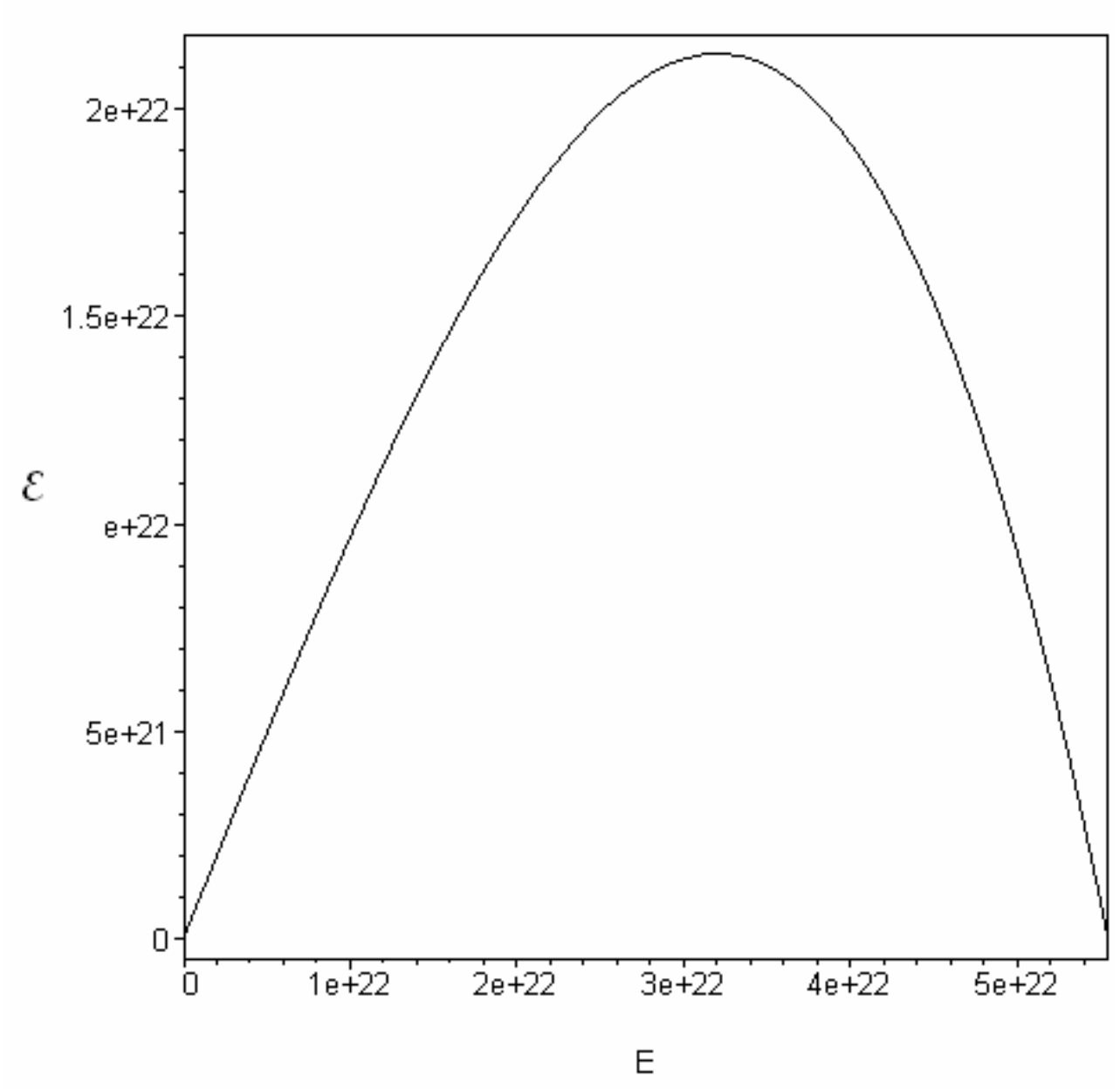

Figure. 1. Emission energy spectrum of the DEMS: at the collapse threshold energy 5.55 $10^{22}$ Mev the DEMS reemission is null.

We will now examine in more detail the behaviour of the DEMS emission energy spectrum (see figure 1).

We can put in evidence three characteristic zones:

1. the first, in Fig. 2, that we call *linear EM zone*, corresponds to the energy interval

$$0 \le E_s < E_{\lim} = 2.03\ 10^{22} \text{ Mev} \quad ;$$

in this zone the massive and EM output of the source is very close to one and the available emitted energy is from the experimental apparatus point of view compatible with the original $E_s$ value;

2. we call the second zone, *critical* or *massive zone*, see Fig. 3; it corresponds to the energy interval,

$$2.03\ 10^{22} \text{ Mev} \le E_s < E_{crit} = 3.2\ 10^{22} \text{ Mev} \ ;$$

in this energy range the output of DEMS decreases appreciably and it achieves the critical output value of 67%;

3. in figure 4 the third and last spectrum zone that we call *collapse* or *break-down zone*, corresponds to the energy interval

$$3.2\ 10^{22} \text{ Mev} \le E_s < E_{col} = 5.55\ 10^{22} \text{ Mev} \ ,$$

in the collapse zone the output decreasing dramatically, dragging the emission to zero.

## 7. Apparently instrumental self-gravitational red shift independence by wavelength

In each matter interaction involving EM fields, if $E$ is a general amount of energy characterising an interaction in which a DEMS with available energy $\mathsf{E}$ is produced, the output (21) allows us to consider the energy loss factor as due to a self-gravitational red shift (SGRS)

$$\eta = \frac{\mathsf{E}}{E} = (1+z)^{-1} \quad . \tag{33.1}$$

As a consequence, each time that an EM interaction within matter is occurring, they transfer an amount of energy $E$ to a local DEMS. The reemission occurs with a content of energy lower than that one predicted by the scattering theory. The difference is due to the SGRS coexisting with the DEMS production

$$\mathsf{E} = E(1+z)^{-1} . \tag{33.2}$$

By using eqs. (22) and (33.1-33.2) it follows that,

$$z = \frac{E}{\mathsf{E}} - 1 = \frac{E^2}{E_{col}^2 - E^2} \quad . \tag{34.1}$$

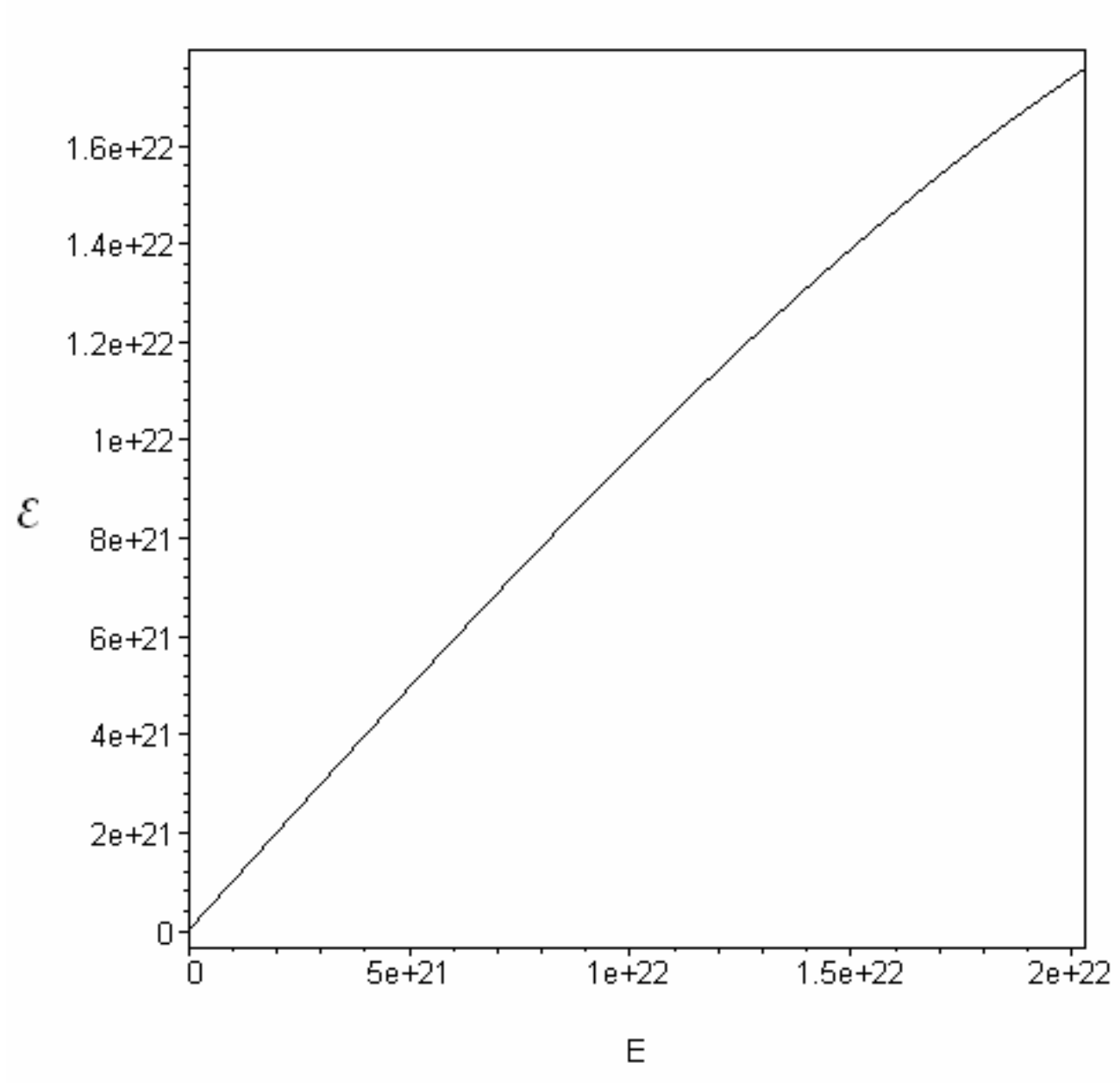

Figure 2. Linear zone of the emission energy spectrum of the DEMS

Since $E_{col} >> E$, the (34.1) can be numerically approximate as

$$z \cong \frac{E^2}{E_{col}^2} \cong 3.24\ 10^{-46}\ E^2 \tag{34.2}$$

In spite of the eq. (34.2) puts in evidence as the self-gravitational red shift for a single DEMS is really dependent by the DEMS energy, instrumentally the SGRS dependence by the wavelength is completely negligible for a very large zone of the DEMS spectrum. In fact, the derivative of eq. (34.2) respect to the wavelength gives

$$dz \cong -10^{-65}\ \lambda^{-3} d\lambda \tag{34.3}$$

And puts in evidence as from the observational point of view a red shift variation is compatible with wavelength independence. Using the (34.3), the absolute variation $|dz|$ is strictly lower than the higher instrumental sensitivity. In this sense, a single SGRS is an experimentally not observable effect because a single drop in energy is negligible when the DEMS energy is lower than the EM-cut one: DEMS and photon can be considered observationally identical; instead the difference could becomes relevant at very high energy of the spectrum or when light photons are along their path statistically subject to a great number of successive scatterings or diffusions process.

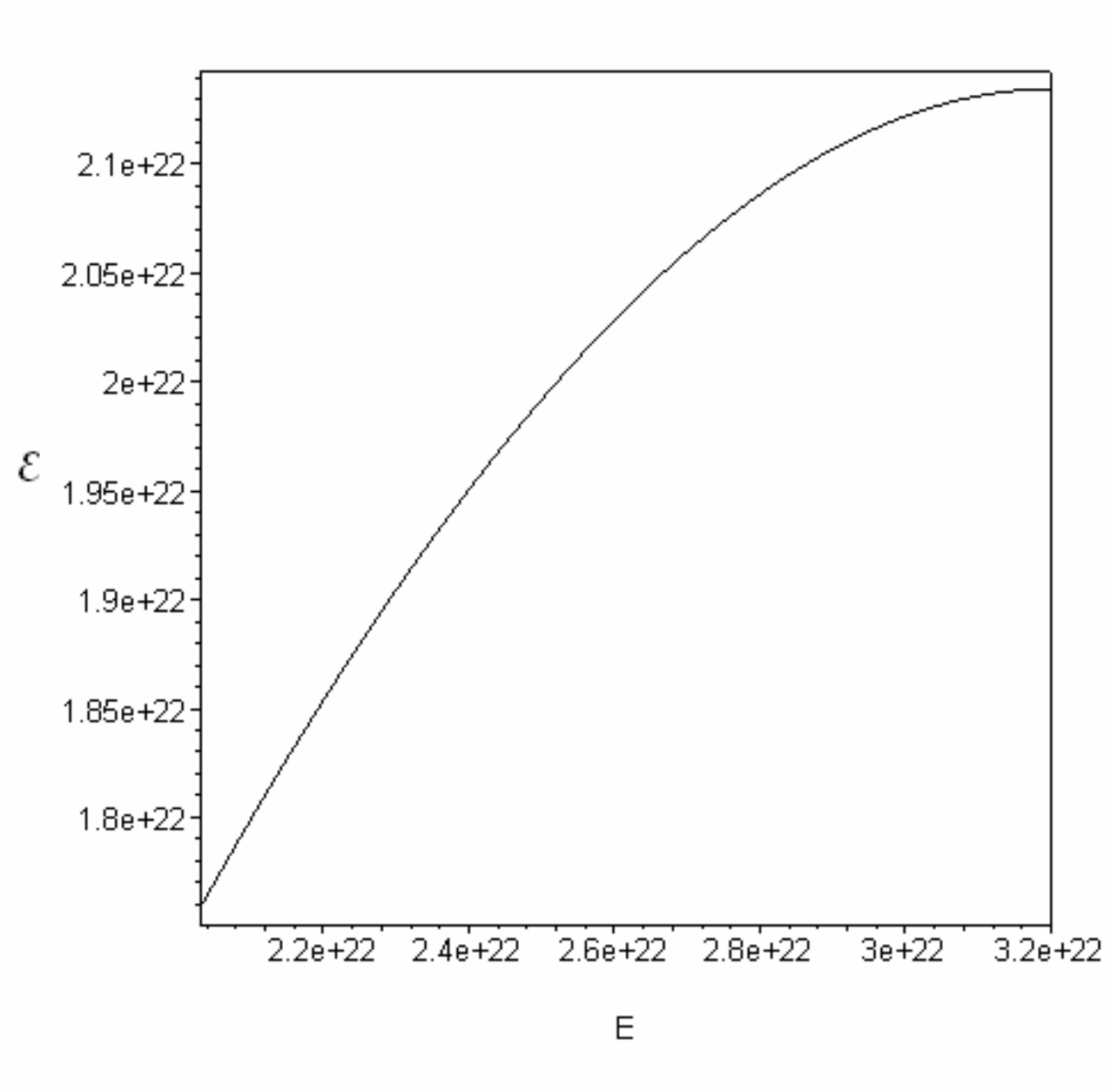

Figure 3. Zone of the maximum of the emission energy spectrum of the DEMS: in the figure the final part of the linear zone and the critical zone are shown.

## 8. Production of Electromagnetic Micro-Black Holes by collapsing DEMSs

When the DEMS energy grows up the collapse threshold, the energy ends to exist under electromagnetic form and only the gravitational field can be observed.

When the energy of the source is greater than the critical value, the DEMS emission converges to zero. The stability of the source is achieved only when the available energy of the source is equal or lower than zero ($E_s \leq 0$); in this case the DEMS acts like a neutral particle with total rest mass

$$M_0 = \hbar\omega/c^2 = 9.89\ 10^{-8}\ \text{Kg} \tag{35.1}$$

and Compton's wavelength

$$\Lambda_0 = 2.23\ 10^{-35}\ \text{m}\,. \tag{35.2}$$

Since the resulting rest mass $M_0$ is very close to Planck's mass and it is originate by the gravitational break-down of a DEMS with integer spin[15], we name this peculiar boson state "*Planck's neutral state",* symbolically $P^0$. Considering the structure of DEMS, the state $P^0$ is associated to a spherical zone of space with maximum external radius $R_{P_0}$ lower than the Schwarzchild's one for the same mass. In fact, considering that the external radius of the SZ is ¾ of the wavelength of the DEMS, in the case of $P^0$, assuming as wavelength the (35b), the radius is lower than the Schwarzchild one:

$$R_{P_0} = \tfrac{3}{4}\Lambda_0 = 1.67\ 10^{-35} < R_{Schwarzchild} = \frac{2\gamma}{c^2}\mathsf{M}_0 = 1.47\ 10^{-34}\,. \tag{36}$$

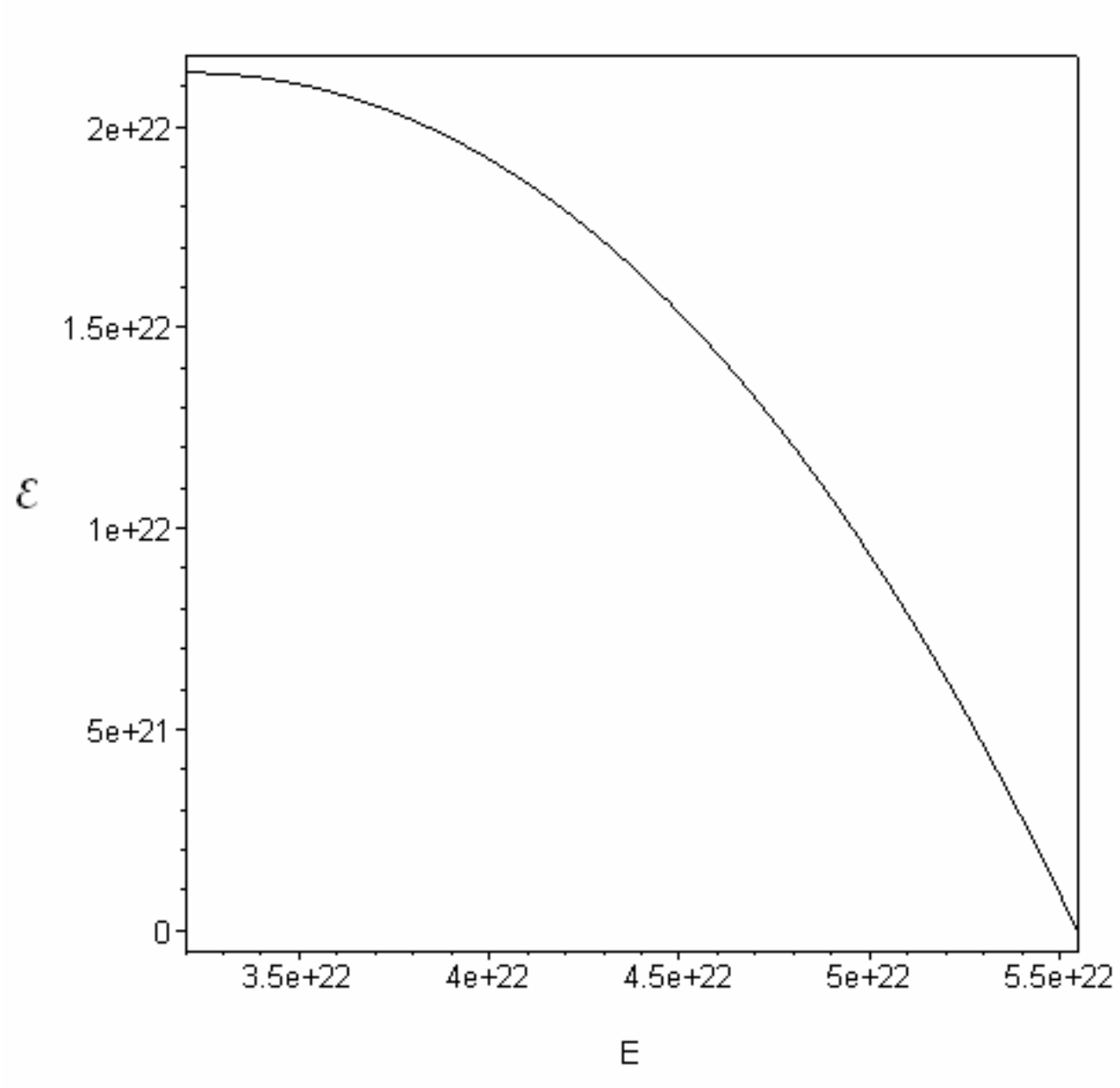

Figure 4. Zone of the maximum of the emission energy spectrum of the DEMS: in the figure the collapse zone is shown.

Eq. (36) puts in evidence that the size of the Schwarzchild's radius is 17.6 times greater than the external radius of the SZ of the corresponding DEMS and 13.2 times greater than the first wave front of the source. This consideration evidence that $P^0$ is a particle state collapsed under the events horizon, i.e. the DEMS becomes a micro-black hole (MBH).

## 9. Conclusion

The SGRS effect appears connected with the production of a sort of "inertia" existing inside the DEMS when the EM mass density grows. As a consequence, all the photons emitted by DEMSs carry an energy lower than that one of the DEMS from which they are produced, i.e. the energy $\mathsf{E} = \hbar\omega'$ of a photon is every only a bit lower then the DEMS one. In this sense, the observer is not able to measure the true energy value of the DEMS, because it is bigger than the photon one of a factor $(1 + z)$, i.e.

$$E = \hbar\omega = \hbar\omega'(1+z)\,.$$

Because the SGRS is a very faint effect, from the point of view of an observer place in the lab frame, the existence of the SGRS effect could not vary too much the local vision of the near Universe, but could vary the physical image of the far Universe. In fact, only when light is subject to a great number of successive interactions with the interstellar medium, the gravitational red shift becomes instrumentally measurable, i.e. only in ancient light an intrinsic gravitational red shift varying along the different line of view could be measured. The effect could produce an extinction of the energy carried by photons as a function of the amount of matter crossed along the line of view of the observer, i.e. as a function of the linear density along the light path. Consequently anisotropies in the measured red shift of the EM background of the far Universe could be expected.

Since does not so much ease to discriminate the anisotropy produced by the SGRS effect from the natural red shift variability of the cosmic background, in order to consider other observable consequences of the SGRS, we suggest that the DEMS available energy for a massive particle emission cannot be greater than the upper bound (29). In fact, assuming that each actual pair of particles in the Universe has been produced by a chain of events in which the EM interaction is mediate by DEMSs, $2.14\ 10^{22}\ \mathrm{Mev}$ represents the top energy for the past, present and future interactions between free charged particles in the Universe. As a consequence, no interaction between pairs can localise enough energy to reach the collapse threshold for the MBH production, than, if Planck's states $P_0$ are existing, they could be originating in the early phase of the Universe or in each other epoch during very high energy cosmic events. If $P_0$ states have been produced in the early phase of the Universe, other phenomena could be correlate with their presence. In fact, the abundance of MBHs in the early Universe could have accelerated the formation of the primordial galaxies, contributing to uniform with their gravitational action, the apparent mass distribution of all galaxies. In this way the states $P_0$ are candidate to be considered an effective component of the "dark matter", contributing to produce on large scale an anomalous galactic discs rotations; moreover the presence of a great amount of MBHs, if uniformly distributed internally and externally to the galaxies, could favourite the gravitational stability observed in the galaxy clusters.